\documentstyle[prd,aps,twocolumn]{revtex}

\begin{document}

\title{
{\normalsize\begin{flushright}
MPI-MIS-23/2001\\
AEI-2001-039\\
hep-th/0105010\\[1.5ex]
\end{flushright}}
Gauge theories of spacetime symmetries
}
\author{
Friedemann Brandt}

\address{
Max-Planck-Institut f\"ur Mathematik in den Naturwissenschaften,\\ 
Inselstra\ss e 22-26, D-04103 Leipzig, Germany;\\
Max-Planck-Institut f\"ur Gravitationsphysik
(Albert-Einstein-In\-sti\-tut),\\ 
Am M\"uhlenberg 1, D-14476 Golm, Germany
\\[1.5ex]
\begin{minipage}{14cm}\rm\quad
Gauge theories of conformal spacetime symmetries are
presented which merge features of Yang-Mills theory and general 
relativity in a new way. The models are local but 
nonpolynomial in the gauge fields, with a nonpolynomial structure 
that can be elegantly written in terms of a metric (or vielbein) 
composed of the gauge fields. General relativity itself emerges from 
the construction as a gauge theory of spacetime translations.
The role of the models within a general classification
of consistent interactions of gauge fields is discussed as well.
\end{minipage}
}

\maketitle

\section{Introduction and conclusion}\label{intro}

In this work new gauge theories of conformal
spacetime symmetries are constructed which merge
features of Yang-Mills theories and 
general relativity in an interesting way. This
concerns both the Lagrangians and the gauge transformations
of these models.
The Lagrangians are local but nonpolynomial in the gauge fields,
as general relativistic Lagrangians
are local but nonpolynomial in the gravitational (metric or vielbein) fields.
In fact, they are formally very similar to general 
relativistic Lagrangians,
except that the metric and vielbein are
polynomials in the conformal gauge fields,
cf.\ eqs.\ (\ref{22}), (\ref{36}). 
Moreover, general relativity
itself emerges from the construction as a 
gauge theory of spacetime translations (see section \ref{GR}).

The (infinitesimal) conformal gauge transformations contain
a Yang-Mills type transformation and a general coordinate transformation,
with the remarkable property that both parts are tied
to each other by the fact that they involve the
{\em same} gauge parameter fields, cf.\ eq.\ (\ref{18b}). 
This unites the symmetry principles 
of Yang-Mills theory and general relativity in an
interesting way and reflects that 
the models are gauge
theories of spacetime symmetries
in a very direct sense. The latter also manifests itself in the 
explicit dependence of
the Lagrangians and the gauge transformations
on conformal spacetime Killing vector fields. 
This, among other things, distinguishes the
models presented here from
gauge theories of conformal symmetries
constructed in the past, such as supergravity models
\cite{Cbf,Mjt,Mej,Kpa,Cih,Fij,Krk,Cbr,Knz,Kea,Dnr,Saj,dtn}, or, 
more recently, models presented in Refs.\ \cite{Wea,Pei}.

At this point a comment seems to be in order.
Particular models constructed in this work admit field
redefinitions (of fields that occur in the action, and of 
gauge parameter fields) which completely remove the explicit dependence of
the Lagrangian and gauge transformations
on conformal Killing vector fields, and cast
the models in more conventional form.
In particular, the standard formulation of general relativity
arises in this way through field redefinitions which trade
metric or vielbein variables for gauge fields of translations. It
is possible, and quite likely, that
the (nonsupersymmetric version of) models constructed in 
\cite{Cbf,Mjt,Mej,Kpa,Cih,Fij,Krk,Cbr,Knz,Kea,Dnr,Saj,dtn}
can be reproduced analogously.
However, it seems to be impossible to eliminate
the dependence on conformal Killing vector fields
in a generic model constructed here. 

The models are not only interesting for their own sake,
but also in the context of a systematic classification of
consistent interactions 
of gauge fields in general, which is quite 
a challenging problem and partly motivated this work.
Such a classification was started in 
\cite{Barnich:1995mt,Barnich:2000zw}
using the BRST cohomological approach to consistent deformations 
of gauge theories \cite{Barnich:1993vg}. 
The starting point of that investigation was the free
Maxwell Lagrangian $L^{(0)}=-(1/4)\sum_A 
F^A_{\mu\nu}F^{\mu\nu A}$ for
a set of vector gauge fields $A_\mu^A$
in flat spacetime. In the deformation approach
one asks whether the action and its gauge symmetries
can be nontrivially deformed, using an expansion in deformation
parameters. 

In \cite{Barnich:1995mt,Barnich:2000zw} 
complete results were derived 
for Poincar\'e invariant deformations of the free Maxwell Lagrangian
to first order in the deformation parameters.
The result is that 
the most general first order deformation which is invariant
under the standard Poincar\'e transformations contains
at most four types of nontrivial interaction vertices:
(i) polynomials in the field strengths
and their first or higher order derivatives;
(ii) Chern-Simons vertices of the form $A\wedge F\wedge\dots\wedge F$
(present only in odd spacetime dimensions);
(iii) cubic interaction vertices $f_{ABC}A^A_\mu A^B_\nu F^{\mu\nu C}$
where $f_{ABC}=f_{[ABC]}$ are antisymmetric constant
coefficients;
(iv) vertices of the form $A_\mu j^\mu$ where
$j^\mu$ is a gauge invariant Noether current of
the free theory\footnote{Note the difference from
vertices (iii): the latter are also of the form $A^A_\mu j^\mu_A$,
but the currents $j^\mu_A=f_{ABC}A^B_\nu F^{\mu\nu C}$
are not gauge invariant.}.
First order deformations 
which are not required to be Poincar\'e invariant 
were also investigated. The results are similar, apart from 
a few (partly unsettled) details
(cf.\ comments
at the end of section 13.2 in \cite{Barnich:2000zw}).

Self-interacting theories for vector gauge fields 
with interaction vertices
(i), (ii) or (iii) are very well known. Those
of type (i) occur, for instance, 
in the Euler-Heisenberg Lagrangian \cite{Heisenberg:1936qt} or
the Born-Infeld theory \cite{Born:1934gh}. Lately,
vertices (i) which are not Lorentz invariant attracted
attention in the context of so-called noncommutative
$U(1)$ gauge theory because the interactions in that model
can be written as an infinite sum of such vertices by means of
a field redefinition (``Seiberg-Witten map'') 
\cite{Seiberg:1999vs}
(field redefinitions of this type are automatically taken care of
by the BRST cohomological approach:
two deformations related by such a field redefinition are
equivalent in that approach).
From the deformation point of view,
vertices (i) and (ii) are somewhat
less interesting because
they are gauge invariant [in case (ii) modulo a total derivative]
under the original gauge transformations of Maxwell theory.

In contrast, vertices (iii) and (iv) are not gauge invariant
under the gauge transformations of the free model; rather, they are
invariant only on-shell (in the free model) modulo a total derivative and
therefore they give rise to nontrivial deformations of the
gauge transformations. This makes them particularly
interesting. Interaction vertices (iii) are of course well known: they
are encountered in Yang-Mills theories \cite{Yang:1954ek,Utiyama:1956sy} 
and lead to a
non-Abelian deformation of the commutator algebra of gauge transformations.
But what about vertices (iv)? Such vertices are
familiar from the coupling of vector gauge
fields to matter fields, such as the coupling of the
electromagnetic gauge field $A_\mu$
to a fermion current $j^\mu=\bar \psi\gamma^\mu\psi$,
but what do we know about vertices involving 
gauge invariant currents made up of the
gauge fields themselves?

As a matter of fact,
it depends on the spacetime dimension whether
or not Poincar\'e invariant
vertices (iv) are present at all. In three dimensions
such vertices exist and occur in 3-dimensional
Freedman-Townsend models \cite{Freedman:1981us,Anco:1995wt}.
In contrast, they do not exist in four dimensions
because Maxwell theory in four dimensions has no 
symmetry that gives rise to a Noether current
needed for a Poincar\'e invariant
vertex (iv) (this follows from the results of \cite{Torre:1995kb}). 
It is likely,
though not proved, that this result in four dimensions
extends to higher dimensions.

However, it must be kept in mind that this result on
vertices (iv) in four dimensions
concerns only Poincar\'e invariant interactions.
The new gauge theories constructed here contain vertices (iv)
that are {\em not} invariant under the standard Poincar\'e transformations
because they involve gauge invariant Noether currents of 
spacetime symmetries themselves. Such vertices exist in all
spacetime dimensions because there is a gauge invariant
form of the Noether currents
of the Poincar\'e symmetries \cite{Jackiw:1978ar,Barnich:1995cq}.
The corresponding deformations of the gauge transformations
incorporate Poincar\'e symmetries in the deformed gauge transformations.
This promotes global Poincar\'e symmetries to local ones, yielding
gauge theories of Poincar\'e symmetries.
In four-dimensional spacetime, the construction can be extended
to the remaining conformal transformations 
because dilatations and special conformal transformations also
give rise to vertices (iv).\footnote{There are infinitely many 
additional vertices (iv) 
that are not Poincar\'e invariant because free Maxwell theory has
infinitely many inequivalent Noether currents 
\cite{Lipkin,Morgan,Kibble,OConnell}. They are not
related to spacetime symmetries. I did not investigate whether
or not they also give rise to interesting gauge theories.}
For this reason I shall focus on models in four-dimensional
spacetime; however, all formulas are also valid in all other dimensions
when restricted to gauge theories
of Poincar\'e symmetries, see section \ref{GR}.

The organization of the paper is the following.
Section \ref{proto}
treats a relatively simple example with only one gauge field and
one vertex (iv) involving a Noether current
of a conformal symmetry in four-dimensional spacetime.
This results in a prototype model with just
one conformal gauge symmetry. 
In section \ref{reform}
the prototype model is rewritten by casting 
its gauge transformations in a more suitable form and introducing
a gauge field dependent ``metric''.
This paves the road for the generalization of the prototype model
in  section \ref{nonabelian} where four-dimensional
gauge theories of the full conformal
algebra or any of its subalgebras are constructed. 
These models involve not only
first order interaction vertices (iv) but in addition
also Yang-Mills type interaction vertices (iii) because
in general 
the involved conformal symmetries do not commute.
Then, in section \ref{matter}, the
construction is further extended by including other
fields (matter fields and gauge fields).
Section \ref{GR} explains the
relation to general relativity.

\section{Prototype model}\label{proto}
 
Let us first examine deformations of the Maxwell action for only
one gauge field $A_\mu$,
\begin{equation}
S^{(0)}=-\frac 14 \int d^4x\,F_{\mu\nu}F^{\mu\nu}
\label{1}
\end{equation}
where $F_{\mu\nu}=\partial_\mu A_\nu-\partial_\nu A_\mu$
is the standard Abelian field strength and
indices $\mu$ are raised with the Minkowski
metric $\eta^{\mu\nu}=\mbox{diag}(+,-,-,-)$
[$F^{\mu\nu}=\eta^{\mu\rho}\eta^{\nu\sigma}F_{\rho\sigma}$].
Action (\ref{1}) is invariant under the gauge transformations
\begin{equation}
\delta^{(0)}_{\lambda} A_\mu=\partial_\mu\lambda
\label{2}
\end{equation}
and under global
conformal transformations
\begin{equation}
\delta_\xi A_\mu=\xi^\nu F_{\nu\mu}
\label{3}
\end{equation}
where $\xi^\mu$ is a conformal Killing vector
field (no matter which one) of flat four-dimensional spacetime,%
\footnote{The construction
is not restricted to flat spacetime but applies analogously to
any fixed background metric $\hat g_{\mu\nu}$ with at least
one conformal Killing vector field $\xi^\mu$. Then
(\ref{4}) turns into 
${\cal L}_\xi \hat g_{\mu\nu} =(1/2) \hat g_{\mu\nu}\hat D_\rho \xi^\rho$
and subsequent formulas change accordingly.}
\begin{equation}
\partial_\mu \xi_\nu+\partial_\nu \xi_\mu=
\frac 12 \eta_{\mu\nu}\partial_\rho \xi^\rho
\quad (\xi_\mu=\eta_{\mu\nu}\xi^\nu).
\label{4}
\end{equation}
(\ref{3}) is the gauge covariant form \cite{Jackiw:1978ar,Barnich:1995cq}
of a conformal transformation and gives rise to the
gauge invariant Noether current
\begin{equation}
j^\mu=\xi^\nu {T_\nu}^\mu,\quad
{T_\nu}^\mu=
-\frac 14\delta_\nu^\mu F_{\rho\sigma}F^{\rho\sigma}
+F_{\nu\rho}F^{\mu\rho}.
\label{5}
\end{equation}
A first order deformation $S^{(1)}$ of action
(\ref{1}) that is of type (iv) and
the corresponding first order deformation $\delta^{(1)}_{\lambda}$
of the gauge transformations (\ref{2}) are
\begin{equation}
S^{(1)}=\int d^4x\, A_\mu j^\mu,\quad
\delta^{(1)}_{\lambda} A_\mu=\lambda\xi^\nu F_{\nu\mu}.
\label{S1}
\end{equation}
Indeed, it can be readily checked that $S^{(1)}$ and
$\delta^{(1)}_{\lambda}$ fulfill the first order
invariance condition
\[
\delta^{(0)}_\lambda S^{(1)}+\delta^{(1)}_{\lambda} S^{(0)}=0.
\]
One may now proceed to higher orders. This amounts to looking
for higher order terms
$S^{(k)}$ and $\delta^{(k)}_{\lambda}$ satisfying
\[
\sum_{i=0}^k\delta^{(i)}_\lambda S^{(k-i)}=0,\quad k=2,3,\dots
\]
It turns out that the deformation exists to all orders but that one
obtains infinitely many terms giving rise to a nonpolynomial
structure. This calls for a more efficient construction
of the complete deformation.
Let me briefly sketch two strategies, without going
into details.
The first one is a detour to a first order formulation:
one casts the original
free Lagrangian in first order form
$(1/4) G^{\mu\nu}(G_{\mu\nu}-2F_{\mu\nu})$ where $G_{\mu\nu}=-G_{\nu\mu}$
are auxiliary fields, deforms this first order model
analogously to (\ref{S1}),
and finally eliminates the auxiliary fields.
Another strategy is the use of a technique applied in
\cite{Brandt:1997ws,Brandt:2000xk}: in view of (\ref{3}),
one defines
a modified field strength $\hat F_{\mu\nu}$ implicitly through
the relations
$\hat F_{\mu\nu}=D_\mu A_\nu-D_\nu A_\mu$ and
$D_\mu A_\nu=\partial_\mu A_\nu-A_\mu \xi^\rho \hat F_{\rho\nu}$, solves
these relations for $\hat F_{\mu\nu}$ and finally constructs the
action and gauge transformations in terms of
$\hat F_{\mu\nu}$ and $A_\mu$. Both
strategies work and yield the same 
action and gauge transformations:
\begin{eqnarray}
&&L=-\frac 14(1+\xi^\rho A_\rho)\hat F_{\mu\nu}\hat F^{\mu\nu},
\label{6}
\\
&&\delta_{\lambda} A_\mu=\partial_\mu{\lambda}
+{\lambda}\,\xi^\nu \hat F_{\nu\mu}
\label{7}
\end{eqnarray}
with $\hat F_{\mu\nu}$ given by
\begin{equation}
\hat F_{\mu\nu}=
F_{\mu\nu}-\frac{A_\mu\xi^\rho F_{\rho\nu}-
A_\nu\xi^\rho F_{\rho\mu}}{1+\xi^\sigma A_\sigma}\ .
\label{8}
\end{equation}
$\hat F_{\mu\nu}$ can be interpreted as the field strength
for the gauge transformations (\ref{7}) because its
gauge transformation does not contain derivatives of $\lambda$:
indeed, a straightforward, though somewhat lengthy, computation gives
\begin{equation}
\delta_{\lambda} \hat F_{\mu\nu}=\frac{\lambda}{1+\xi^\sigma A_\sigma}\,
{\cal L}_\xi \hat F_{\mu\nu}
\label{7'}
\end{equation}
where ${\cal L}_\xi$ is the standard Lie derivative along
$\xi^\mu$, 
\[
{\cal L}_\xi\hat F_{\mu\nu}=\xi^\rho\partial_\rho \hat F_{\mu\nu}
+\partial_\mu\xi^\rho \hat F_{\rho\nu}
+\partial_\nu\xi^\rho \hat F_{\mu\rho}\ .
\]
Using (\ref{7'}), as well as (\ref{4}), it is easy to verify that 
the Lagrangian (\ref{6}) transforms under the
gauge transformations (\ref{7}) into a total derivative,
\begin{equation}
\delta_{\lambda} L=-\frac 14\partial_\mu(
\lambda\,\xi^\mu \hat F_{\rho\nu}\hat F^{\rho\nu}).
\label{9}
\end{equation}
Furthermore, owing to (\ref{7'}),
the algebra of the gauge transformations (\ref{7}) is obviously
Abelian, i.e., two gauge transformations with different parameter
fields, denoted by $\lambda$ and $\lambda'$, respectively, commute:
\begin{equation}
[\delta_{\lambda},\delta_{\lambda'}]=0.
\label{10}
\end{equation}
I remark that, for notational convenience, I have suppressed
the gauge coupling constant (= deformation parameter) in the
formulas given above; it can be easily introduced in the usual
way by substituting rescaled fields $\kappa A_\mu$ and
$\kappa\lambda$ for $A_\mu$ and $\lambda$,
respectively, and then dividing the Lagrangian by $\kappa^2$.
Expanding
the resulting action and gauge transformations in $\kappa$,
one obtains
$S=S^{(0)}+\kappa S^{(1)}+O(\kappa^2)$ and 
$\delta_\lambda=\delta_\lambda^{(0)}
+\kappa\delta_\lambda^{(1)}+O(\kappa^2)$ with 
$S^{(1)}$ and $\delta_\lambda^{(1)}$
as in (\ref{S1}). This shows that (\ref{6}) and (\ref{7})
complete
the first order deformation (\ref{S1})
to all orders. Note that the
completion contains infinitely many terms and is nonpolynomial
but local in the gauge fields, as promised.

\section{Reformulation of the prototype model}\label{reform}

In the remainder of this work
I shall first rewrite and then
generalize the prototype model with the Lagrangian
(\ref{6}) and the gauge transformations (\ref{7}).
A surprising feature of the Lagrangian
(\ref{6}) is that its nonpolynomial structure
can be written in terms of the ``metric''
\begin{equation}
g_{\mu\nu}=\eta_{\mu\nu}+\xi_\mu A_\nu+\xi_\nu A_\mu
+\xi_\rho\xi^\rho A_\mu A_\nu\ ,
\label{11}
\end{equation}
where, again, $\xi_\mu=\eta_{\mu\nu}\xi^\nu$. The
inverse and determinant of this metric are
\begin{eqnarray*}
&&g^{\mu\nu}=\eta^{\mu\nu}-
\frac{\xi^\mu A^\nu+\xi^\nu A^\mu}{1+\xi^\sigma A_\sigma}
+\frac{A_\rho A^\rho \xi^\mu \xi^\nu}{(1+\xi^\sigma A_\sigma)^2}\ ,
\\
&&\det(g_{\mu\nu})=-(1+\xi^\mu A_\mu)^2,
\end{eqnarray*}
where $A^\mu=\eta^{\mu\nu}A_\nu$.
Using these formulas one readily verifies that the Lagrangian (\ref{6})
can be written as
\begin{equation}
L=-\frac 14 \sqrt{g}g^{\mu\rho}g^{\nu\sigma}F_{\mu\nu}F_{\rho\sigma}
\label{12}
\end{equation} 
where $F_{\mu\nu}=\partial_\mu A_\nu-\partial_\nu A_\mu$ and 
$\sqrt g=|\det(g_{\mu\nu})|^{1/2}=1+\xi^\mu A_\mu$ (assuming
$1+\xi^\mu A_\mu>0$).
Furthermore, it can be easily checked that 
the gauge transformations (\ref{7}) can be rewritten as
\begin{equation}
\delta_{\omega}A_\mu=\partial_\mu\omega+
\omega\xi^\nu\partial_\nu A_\mu+\partial_\mu(\omega\xi^\nu) A_\nu
\label{13}
\end{equation}
where $\omega$ is constructed of 
$\lambda$, $\xi^\mu$ and $A_\mu$ according to
\begin{equation}
\omega=\frac{\lambda}{1+\xi^\mu A_\mu}\ .
\label{14}
\end{equation}
(\ref{13}) is exactly
the same transformation as (\ref{7}), but written
in terms of $\omega$ instead of
$\lambda$. Since $\lambda$ was completely arbitrary,
$\omega$ is also completely arbitrary, and can thus
be used as gauge parameter field in place of $\lambda$.
Note that (\ref{13}) is polynomial in the gauge fields, in
contrast to (\ref{7}).

To understand the gauge invariance of the model, and to
generalize it subsequently,
the following observation is crucial: under 
the gauge transformations
(\ref{13}) of the gauge fields, the metric
(\ref{11}) transforms according to
\begin{equation}
\delta_{\omega}g_{\mu\nu}={\cal L}_\varepsilon g_{\mu\nu}
-\frac 12 g_{\mu\nu}\omega \partial_\rho\xi^\rho,
\label{15}
\end{equation}
where ${\cal L}_\varepsilon g_{\mu\nu}$ is the Lie derivative of $g_{\mu\nu}$
along $\varepsilon^\mu=\omega\xi^\mu$:
\begin{eqnarray}
&
{\cal L}_\varepsilon g_{\mu\nu}=\varepsilon^\rho\partial_\rho g_{\mu\nu}
+\partial_\mu\varepsilon^\rho g_{\rho\nu}
+\partial_\nu\varepsilon^\rho g_{\mu\rho},
&
\nonumber\\
&
\varepsilon^\mu=\omega\xi^\mu.
&
\label{15b}
\end{eqnarray}
In order to
verify equation (\ref{15}), one has to use the 
conformal Killing vector equations (\ref{4}). 
Equations (\ref{13}) and (\ref{15})
make it now easy to understand the gauge invariance of the action with
Lagrangian (\ref{12}). 
Note that the last two terms on the right hand side of
(\ref{13}) are nothing but the Lie derivative
${\cal L}_\varepsilon A_\mu$
of $A_\mu$ along $\varepsilon^\mu$:
\[
\delta_{\omega}A_\mu=\partial_\mu\omega+{\cal L}_\varepsilon A_\mu.
\]
Hence the gauge transformation
of $A_\mu$ is the sum of a standard Abelian
gauge transformation with parameter $\omega$ and
a general coordinate transformation with parameters
$\varepsilon^\mu$ [of course, 
these two transformations are related because of 
$\varepsilon^\mu=\omega\xi^\mu$].
As a consequence, the gauge transformation of $F_{\mu\nu}$
is given just by the Lie derivative
along $\varepsilon^\mu$,
$\delta_{\omega}F_{\mu\nu}={\cal L}_\varepsilon F_{\mu\nu}$.
(\ref{15}) has the form
of a general coordinate transformation of $g_{\mu\nu}$ with
parameters $\varepsilon^\mu$ plus a Weyl transformation with
parameter $-(1/2)\omega \partial_\rho\xi^\rho$.
As the Lagrangian is invariant under Weyl transformations
of $g_{\mu\nu}$
(we are still discussing the four-dimensional
case), it transforms under 
gauge transformations $\delta_{\omega}$ just like a scalar
density under general coordinate transformations
with parameters $\varepsilon^\mu$: 
$\delta_{\omega}L=\partial_\mu (\varepsilon^\mu L)$.
This is exactly equation (\ref{9}), owing
to $\varepsilon^\mu=\omega\xi^\mu=\lambda\xi^\mu/(1+\xi^\nu A_\nu)$
and $L/(1+\xi^\nu A_\nu)=-(1/4)\hat F_{\mu\nu}\hat F^{\mu\nu}$.
A final remark on the prototype model is that the 
gauge transformations no longer commute when
expressed in terms of $\omega$
rather than in terms of $\lambda$:
\begin{equation}
[\delta_{\omega},\delta_{\omega'}]=\delta_{\omega''} ,\quad
\omega''=
\omega'\xi^\mu\partial_\mu\omega-\omega\xi^\mu\partial_\mu\omega'.
\label{16}
\end{equation}
The reason for this is that the redefinition (\ref{14}) of the
gauge parameter field involves the gauge field $A_\mu$.

\section{Generalization}\label{nonabelian}

The prototype model found above will now be generalized
by gauging more than only
one conformal symmetry in four-dimensional flat spacetime.
Let ${\cal G}$ be the Lie algebra of the full
conformal group or any of its subalgebras. Let
us pick a basis of ${\cal G}$ and label its elements by
an index $A$ [since the conformal group in
four dimensions is 15-dimensional, we have $A=1,\dots,N$
with $1\leq N\leq 15$]. The corresponding set of
conformal Killing vector fields is denoted by
$\{\xi^\mu_A\}$. Since ${\cal G}$ is a Lie algebra, one can
choose the $\xi$'s such that
\begin{equation}
\xi_A^\nu \partial_\nu\xi_B^\mu-\xi_B^\nu \partial_\nu\xi_A^\mu
={f_{BA}}^C\xi_C^\mu
\label{17}
\end{equation}
where ${f_{AB}}^C$ are the structure constants of
${\cal G}$ in the chosen basis. I associate
one gauge field $A_\mu^A$ and one gauge parameter field $\omega^A$
with each element of ${\cal G}$ and introduce the following
generalization of the gauge transformations
(\ref{13}):
\begin{equation}
\delta_{\omega} A_\mu^A=D_\mu\omega^A+
\omega^B\xi^\nu_B\partial_\nu A_\mu^A+\partial_\mu(\omega^B\xi^\nu_B)A_\nu^A
\label{18}
\end{equation}
where 
\begin{equation}
D_\mu\omega^A=\partial_\mu \omega^A+A_\mu^B{f_{BC}}^A\omega^C.
\label{19}
\end{equation}
The part $D_\mu\omega^A$ of $\delta_{\omega} A_\mu^A$
is familiar from Yang-Mills theory; the remaining part
is the Lie derivative
of $A_\mu^A$ along a vector field $\varepsilon^\mu$ containing
the gauge parameter fields $\omega^A$,
\begin{equation}
\delta_{\omega} A_\mu^A=D_\mu\omega^A+{\cal L}_\varepsilon A_\mu^A,\quad
\varepsilon^\mu=\omega^B\xi^\mu_B.
\label{18b}
\end{equation}
The commutator of two gauge transformations is
\begin{eqnarray}
&[\delta_{\omega},\delta_{\omega'}]=\delta_{\omega''}\ ,&
\nonumber\\
&\omega^{\prime\prime A}=\omega^{B}\omega^{\prime C}{f_{BC}}^A
+\omega^{\prime B}\xi_B^\mu\partial_\mu \omega^A
-\omega^{B}\xi_B^\mu\partial_\mu \omega^{\prime A}.&
\label{18a}
\end{eqnarray} 
The crucial step for constructing an action which is
invariant under these gauge transformations
is the following generalization of the prototype metric (\ref{11}):
\begin{equation}
g_{\mu\nu}=\eta_{\mu\nu}+\xi_{A\mu} A^A_\nu+\xi_{A\nu} A^A_\mu
+\xi_{A\rho}\xi_B^\rho A^A_\mu A^B_\nu,
\label{22}
\end{equation}
with $\xi_{A\mu}=\eta_{\mu\nu}\xi^\nu_A$.
This metric behaves under gauge transformations (\ref{18})
similarly as the prototype metric (\ref{11}) under
gauge transformations (\ref{13}):
\begin{equation}
\delta_{\omega}g_{\mu\nu}={\cal L}_\varepsilon g_{\mu\nu}
-\frac 12 g_{\mu\nu}\omega^A \partial_\rho\xi^\rho_A,
\label{23}
\end{equation}
with $\varepsilon^\mu$ as in (\ref{18b}).
To verify (\ref{23}), one has to use 
(\ref{4}) (which holds for each $\xi_A^\mu$)
and (\ref{17}). Note that (\ref{18}) is the sum
of a Yang-Mills gauge transformation with parameter fields
$\omega^B$ and a general coordinate
transformation with parameter fields $\varepsilon^\mu=\omega^B\xi^\mu_B$, 
while (\ref{23}) has the form of a general coordinate
transformation with parameters $\varepsilon^\mu$ plus a Weyl transformation
with parameter $-(1/2)\omega^A \partial_\rho\xi^\rho_A$.
This immediately implies that the following Lagrangian
is invariant modulo a total derivative
under gauge transformations (\ref{18}):
\begin{equation}
L=-\frac 14 \sqrt{g}g^{\mu\rho}g^{\nu\sigma}F^A_{\mu\nu}F^B_{\rho\sigma}
d_{AB},
\label{24}
\end{equation}
where $d_{AB}$ is a symmetric ${\cal G}$-invariant tensor,
\begin{equation}
d_{AB}=d_{BA},\quad {f_{CA}}^D d_{DB}+{f_{CB}}^D d_{AD}=0,
\label{21}
\end{equation}
and the
$F^A_{\mu\nu}$ are field strengths familiar from
Yang-Mills theory:
\begin{equation}
F_{\mu\nu}^A=\partial_\mu A_\nu^A-\partial_\nu A_\mu^A
+{f_{BC}}^A A_\mu^B A_\nu^C.
\label{20}
\end{equation}
Owing to (\ref{21}), the 
Lagrangian (\ref{24}) is invariant under
Yang-Mills transformations of the $F_{\mu\nu}^A$.
Furthermore it is invariant under Weyl transformations of $g_{\mu\nu}$.
Hence, it
transforms under gauge transformations (\ref{18})
just like a scalar density under general coordinate
transformations with parameters $\varepsilon^\mu=\omega^A\xi^\mu_A$:
\begin{equation}
\delta_{\omega}L=\partial_\mu (\omega^A\xi^\mu_A L).
\label{26}
\end{equation}
Again, the Lagrangian is local but nonpolynomial in
the gauge fields because it contains the inverse 
metric $g^{\mu\nu}$. The latter is
\begin{equation}
g^{\mu\nu}=\eta^{\mu\nu}-\xi^\mu_A \hat A^{A\nu}
-\xi^\nu_A \hat A^{A\mu}
+\xi^\mu_A \xi^\nu_B \hat A^A_\rho\hat A^{B\rho},
\label{27a}
\end{equation}
where $\hat A^{A\mu}=\eta^{\mu\nu}\hat A^A_\nu$, with
\begin{equation}
\hat A^A_\mu= A^B_\mu {E_B}^A,\quad
{E_B}^C(\delta^A_C+\xi^\mu_C A_\mu^A)=\delta^A_B.
\label{27b}
\end{equation}
The second equation in (\ref{27b}) expresses that 
the ${E_B}^A$ are the entries of  
a matrix $E$ which inverts the matrix $1+M$ where $M$
is the matrix
with entries $\xi^\mu_B A_\mu^A$. $E$ can thus be written as an infinite
(geometric) series
of matrix products of $M$:
\begin{equation}
E=\sum_{k=0}^\infty (-M)^k,\quad {M_B}^A\equiv
\xi^\mu_B A_\mu^A.
\label{27c}
\end{equation}
A gauge coupling constant $\kappa$ can be
introduced as before by means of the substitutions 
$A_\mu^A\rightarrow \kappa A_\mu^A$,
$\omega^A\rightarrow \kappa \omega^A$, $L\rightarrow L/\kappa^2$.
Equivalently, one may use 
${f_{AB}}^C\rightarrow \kappa {f_{AB}}^C$, $\xi^\mu_A\rightarrow 
\kappa\xi^\mu_A$.
Of course, the zeroth order Lagrangian is
positive definite only for appropriate choices of ${\cal G}$.
For instance, one may choose a ${\cal G}$ that is Abelian or
compact; then there is a basis of ${\cal G}$ such that $d_{AB}=\delta_{AB}$.
The simplest case is a one-dimensional ${\cal G}$ and
reproduces the prototype model.
Choices such as
${\cal G}=so(2,4)$ (full conformal algebra) or ${\cal G}=so(1,3)$
(Lorentz algebra) do not give a positive definite zeroth order Lagrangian
because these algebras are not compact (one cannot achieve
$d_{AB}=\delta_{AB}$).

\section{Inclusion of matter fields and further gauge fields}\label{matter}

Using the metric (\ref{22}), it is straightforward
to extend the models of the
previous section so as to include
further fields.
First I discuss the case of just one (real) scalar field $\phi$
and introduce the gauge transformation
\begin{equation}
\delta_\omega\phi=
\omega^A\xi^\mu_A\partial_\mu\phi
+\frac 14 \phi\omega^A\partial_\mu\xi^\mu_A.
\label{30}
\end{equation}
A contribution to the Lagrangian which is gauge
invariant modulo a total derivative is
\begin{equation}
L_\phi=
\frac 12 \sqrt g g^{\mu\nu}\partial_\mu \phi \partial_\nu \phi
-\frac 1{12}\sqrt g R \phi^2
\label{33}
\end{equation}
with $g_{\mu\nu}$ and $g^{\mu\nu}$ as before in
(\ref{22}) and (\ref{27a}), and $R$ the Riemannian curvature scalar
built from $g_{\mu\nu}$,
\begin{eqnarray*}
R&=&g^{\mu\nu}{R_{\mu\rho\nu}}^\rho,
\\
{R_{\mu\nu\rho}}^\sigma&=&\partial_\mu{\Gamma_{\nu\rho}}^\sigma
+{\Gamma_{\mu\lambda}}^\sigma{\Gamma_{\nu\rho}}^\lambda
-(\mu\leftrightarrow\nu),
\\
{\Gamma_{\mu\nu}}^\rho&=&\frac 12g^{\rho\sigma}
(\partial_\mu g_{\nu\sigma}
+\partial_\nu g_{\mu\sigma}-\partial_\sigma g_{\mu\nu}).
\end{eqnarray*}
Using (\ref{23}), one easily derives
the gauge variation of $R$:
\begin{eqnarray}
\delta_\omega R&=&\varepsilon^\mu \partial_\mu R
+\frac 12 R\omega^A\partial_\mu\xi^\mu_A
\nonumber\\
&&-\frac 32 g^{\mu\nu}(\partial_\mu\partial_\nu
-{\Gamma_{\mu\nu}}^\rho\partial_\rho)
(\omega^A\partial_\sigma\xi^\sigma_A).
\label{34}
\end{eqnarray}
This makes it is easy to verify
the gauge invariance of
(\ref{33}): $L_\phi$ transforms as
a scalar density under standard general coordinate transformations
of $g_{\mu\nu}$ and $\phi$; therefore
the first term in (\ref{30}) and the first term in (\ref{23})
make a contribution $\partial_\mu(\varepsilon^\mu L_\phi)$ to
$\delta_\omega L_\phi$; the second terms in (\ref{30})
and (\ref{23}) contribute a total derivative to
$\delta_\omega L_\phi$ because $L_\phi$ is invariant
modulo a total derivative under Weyl transformations
of $g_{\mu\nu}$ and $\phi$ with 
weights of ratio $-2$ (in four dimensions). The 
complete transformation reads
\begin{equation}
\delta_\omega  L_\phi=\partial_\mu\Big[
\omega^A\xi^\mu_A L_\phi+\frac 18\sqrt gg^{\mu\nu} \phi^2
\partial_\nu(\omega^A\partial_\rho\xi^\rho_A)
\Big].
\label{35}
\end{equation}

To include fermions, I introduce the ``vierbein''
\begin{equation}
{e_\mu}^\nu=\delta^\nu_\mu+\xi^\nu_A A_\mu^A.
\label{36}
\end{equation}
The term vierbein is used because
${e_\mu}^\nu$ is related to the ``metric'' (\ref{22}) through
\begin{equation}
g_{\mu\nu}=\eta_{\rho\sigma} {e_\mu}^\rho  {e_\nu}^\sigma.
\label{37}
\end{equation}
Furthermore the vierbein transforms under the gauge transformations
(\ref{18}) according to
\begin{eqnarray}
\delta_\omega {e_\mu}^\nu&=&
\varepsilon^\rho\partial_\rho {e_\mu}^\nu
+\partial_\mu\varepsilon^\rho {e_\rho}^\nu
\nonumber\\
&&
+{C_\rho}^\nu {e_\mu}^\rho-\frac 14{e_\mu}^\nu\omega^A\partial_\rho\xi^\rho_A
\label{38}
\end{eqnarray}
with $\varepsilon^\mu$ as in (\ref{18b}) and
\begin{equation}
{C_\mu}^\nu=-\frac 12\omega^A(\partial_\mu\xi^\nu_A
-\eta^{\nu\sigma}\eta_{\mu\rho}\partial_\sigma\xi^\rho_A).
\label{39}
\end{equation}
Note that (\ref{38}) has indeed the familiar form
of the transformation of vierbein fields in general relativity:
the lower index of ${e_\mu}^\nu$ transforms
as a ``world index'' (it sees only the general coordinate
transformation with parameters $\varepsilon^\mu$) while the
upper index transforms as a ``Lorentz index'' (it sees
only ``Lorentz transformations with parameters ${C_\mu}^\nu$''
-- the Lorentz character is due to
$C^{\mu\nu}=-C^{\nu\mu}$ where $C^{\mu\nu}=\eta^{\mu\rho}{C_\rho}^\nu$).
In addition (\ref{38})
contains a Weyl transformation with parameter
$-(1/2)\omega^A\partial_\rho\xi^\rho_A$.
I now define a ``spin connection'' ${\omega_\mu}^{\nu\rho}$:
\begin{eqnarray}
{\omega_\mu}^{\nu\rho}&=&
{E_\sigma}^\nu {E_\lambda}^\rho\eta^{\sigma\kappa}\eta^{\lambda\tau}
\omega_{\mu\kappa\tau},
\nonumber\\
\omega_{\mu\nu\rho}&=&
\omega_{[\mu\nu]\rho}-\omega_{[\nu\rho]\mu}+\omega_{[\rho\mu]\nu},
\nonumber\\
\omega_{[\mu\nu]\rho}&=&\frac 12 {e_\rho}^\sigma\eta_{\sigma\lambda}
(\partial_\mu{e_\nu}^\lambda-\partial_\nu{e_\mu}^\lambda)
\end{eqnarray}
where ${E_\mu}^\nu$ is the inverse vierbein (${E_\mu}^\nu{e_\nu}^\rho
=\delta_\mu^\rho$),
\begin{equation}
{E_\mu}^\nu=\delta_\mu^\nu-\hat A^A_\mu \xi^\nu_A
\label{viel}
\end{equation}
with $\hat A^A_\mu$ as in (\ref{27b}). Since ${\omega_\mu}^{\nu\rho}$
is constructed of ${e_\mu}^\nu$ in exactly the same manner as
one constructs the spin connection of the vierbein
in general relativity, one infers from (\ref{38}) that
${\omega_\mu}^{\nu\rho}$ transforms under the gauge transformations
(\ref{18}) according to
\begin{eqnarray}
\delta_\omega {\omega_\mu}^{\nu\rho}&=&
\partial_\mu C^{\nu\rho}-{\omega_\mu}^{\sigma\nu}{C_\sigma}^\rho
+{\omega_\mu}^{\sigma\rho}{C_\sigma}^\nu
\nonumber\\
&&+\varepsilon^\sigma\partial_\sigma {\omega_\mu}^{\nu\rho}
+\partial_\mu \varepsilon^\sigma {\omega_\sigma}^{\nu\rho}
\nonumber\\
&&+\frac 14 ({e_\mu}^\rho{E_\sigma}^\nu-{e_\mu}^\nu{E_\sigma}^\rho)
\eta^{\sigma\lambda}\partial_\lambda
(\omega^A\partial_\tau\xi^\tau_A)
\end{eqnarray}
where $C^{\nu\rho}=\eta^{\nu\sigma}{C_\sigma}^\rho$ with
${C_\sigma}^\rho$ as in (\ref{39}).
I denote a fermion field by $\psi$ (without displaying
its spinor indices), and introduce the gauge
transformations
\begin{equation}
\delta_\omega\psi=
\omega^A\xi^\mu_A\partial_\mu\psi
-\frac 12C^{\mu\nu}\sigma_{\mu\nu}\psi
+\frac 38 \psi\omega^A\partial_\mu\xi^\mu_A
\label{40}
\end{equation}
where $4\sigma_{\mu\nu}$ is the commutator
of $\gamma$-matrices, using the conventions
\begin{eqnarray*}
&\gamma^\mu\gamma^\nu+\gamma^\nu\gamma^\mu=2\eta^{\mu\nu},&
\\
&\sigma_{\mu\nu}=\frac 14(\gamma_\mu\gamma_\nu-\gamma_\nu\gamma_\mu),
\quad \gamma_\mu=\eta_{\mu\nu}\gamma^\nu.&
\end{eqnarray*}
A contribution to the Lagrangian which is invariant modulo
a total derivative under
the gauge transformations (\ref{18}) and (\ref{40}) is
\begin{equation}
L_\psi={\mathrm{i}}\sqrt g\, \bar\psi\gamma^\nu {E_\nu}^\mu 
(\partial_\mu\psi+\frac 12 {\omega_\mu}^{\sigma\rho}\sigma_{\sigma\rho}\psi).
\label{41}
\end{equation}
$L_\psi$ transforms under the gauge transformations
like a scalar density under general coordinate transformations
with parameters $\varepsilon^\mu=\omega^A\xi^\mu_A$ because the
``Lorentz'' and
``Weyl'' parts of the gauge transformation of the fermion,
vierbein and spin connection cancel each other completely,
\begin{equation}
\delta_\omega L_\psi=\partial_\mu(\omega^A\xi^\mu_A L_\psi).
\label{43}
\end{equation}

The inclusion of standard Yang-Mills gauge fields $A_\mu^I$
is even simpler:
the contribution to the Lagrangian is just the
standard Yang-Mills Lagrangian in the metric
(\ref{22}),
\begin{eqnarray}
L_{\mathrm{YM}}&=&
-\frac 14 \sqrt{g}g^{\mu\rho}g^{\nu\sigma}F^I_{\mu\nu}F^J_{\rho\sigma}
d_{IJ},
\label{44}\\
F_{\mu\nu}^I&=&\partial_\mu A_\nu^I-\partial_\nu A_\mu^I
+{f_{JK}}^I A_\mu^J A_\nu^K
\label{45}
\end{eqnarray}
where ${f_{JK}}^I$ and $d_{IJ}$ are the structure constants 
and an invariant symmetric tensor of some
Lie algebra ${\cal G}_{\mathrm{YM}}$. Note that
the difference from (\ref{24}) is that now the field strengths
$F_{\mu\nu}^I$ involve the gauge fields of ${\cal G}_{\mathrm{YM}}$ while
the metric $g_{\mu\nu}$ is composed of the gauge fields
of ${\cal G}$. The conformal gauge transformations of $A_\mu^I$ are just
the standard Lie derivatives along $\varepsilon^\mu=\omega^A\xi_A^\mu$,
\begin{equation}
\delta_{\omega} A_\mu^I=
\omega^B\xi^\nu_B\partial_\nu A_\mu^I+\partial_\mu(\omega^B\xi^\nu_B)A_\nu^I.
\label{46}
\end{equation}
Since
$L_{\mathrm{YM}}$ is invariant under Weyl transformations
of $g_{\mu\nu}$, it transforms under the conformal gauge transformations
(\ref{18}) and (\ref{46}) like a scalar
density under general coordinate transformations with
parameters $\varepsilon^\mu$,
\begin{equation}
\delta_\omega L_{\mathrm{YM}}=\partial_\mu(\omega^A\xi^\mu_A L_{\mathrm{YM}}).
\label{47}
\end{equation}
In addition $L_{\mathrm{YM}}$ is invariant under the usual
Yang-Mills gauge transformations
$\delta_{\alpha} A_\mu^I=
\partial_\mu \alpha^I+A_\mu^J{f_{JK}}^I\alpha^K$
for arbitrary gauge parameter fields $\alpha^I$.

It is straightforward to construct
further interaction terms, such as $\sqrt g\phi^4$ or
Yukawa-interactions $\sqrt g\phi\bar\psi\psi$, and to
extend the construction to scalar fields
or fermions transforming nontrivially under
${\cal G}_{\mathrm{YM}}$. In fact, it is even possible
to construct models where the ``matter fields'' transform
under ${\cal G}$ according to a nontrivial
representation. I shall only discuss the case of scalar fields
transforming under a nontrivial representation of ${\cal G}$;
the extension to fermions is
straightforward. Of course, the notion ``scalar
fields'' should be used cautiously when these
fields sit in a nontrivial representation of ${\cal G}$
as they may or may not
transform nontrivially under Lorentz transformations (depending
on the choice of ${\cal G}$ and its representation).
I denote these ``scalar fields'' by $\phi^i$.
The corresponding  representation matrices of ${\cal G}$
are denoted by $T_A$ and chosen such that they represent
${\cal G}$ with the same structure constants ${f_{AB}}^C$
as in (\ref{17}), i.e.,
\begin{equation}
T_{Ak}^i T_{Bj}^k-T_{Bk}^i T_{Aj}^k={f_{AB}}^C T_{Cj}^i.
\label{28}
\end{equation}
Further properties of the representation will not
matter to the construction.
In place of (\ref{30}),
the gauge transformations now read
\begin{equation}
\delta_\omega\phi^i=
-\omega^A T_{Aj}^i\phi^j+\omega^A\xi^\mu_A\partial_\mu\phi^i
+\frac 14 \phi^i\omega^A\partial_\mu\xi^\mu_A.
\label{30A}
\end{equation}
Accordingly, one introduces covariant derivatives
\begin{equation}
D_\mu\phi^i=\partial_\mu\phi^i+A_\mu^A T_{Aj}^i\phi^j.
\label{31}
\end{equation}
These covariant derivatives transform under gauge transformations
(\ref{18}), (\ref{30A}) according to
\begin{eqnarray*}
\delta_\omega D_\mu\phi^i&=&
-\omega^A T_{Aj}^i D_\mu\phi^j+{\cal L}_\varepsilon D_\mu\phi^i
\nonumber\\
&&+\frac 14 (D_\mu\phi^i)\omega^A\partial_\nu\xi^\nu_A
+\frac 14\phi^i\partial_\mu(\omega^A\partial_\nu\xi^\nu_A)
\end{eqnarray*}
where ${\cal L}_\varepsilon D_\mu\phi^i
=\varepsilon^\nu\partial_\nu D_\mu\phi^i+
\partial_\mu\varepsilon^\nu D_\nu\phi^i$ with $\varepsilon^\mu$ 
as in (\ref{18b}). The generalization of the
Lagrangian (\ref{33}) is simply
\begin{equation}
\tilde{L}_\phi=
\sqrt g \Big[\frac 12 g^{\mu\nu}D_\mu \phi^i D_\nu \phi^j
-\frac 1{12} R \phi^i \phi^j\Big] d_{ij}
\label{33A}
\end{equation}
where $d_{ij}$ is a ${\cal G}$-invariant symmetric tensor,
\begin{equation}
d_{ij}=d_{ji},\quad d_{kj}T_{Ai}^k+d_{ik}T_{Aj}^k=0.
\label{29}
\end{equation}
Using (\ref{29}) and arguments analogous to those that led to (\ref{35}),
one infers that
\[
\delta_\omega  \tilde{L}_\phi=\partial_\mu\Big[
\omega^A\xi^\mu_A L_\phi+\frac 18\sqrt gg^{\mu\nu} \phi^i\phi^jd_{ij}
\partial_\nu(\omega^A\partial_\rho\xi^\rho_A)
\Big].
\]

\section{Relation to general relativity}\label{GR}

So far we have worked in four-dimensional spacetime.
Actually the whole construction goes through
without any change
in an arbitrary dimension if we restrict it to
isometries of the flat metric rather than 
considering all conformal
symmetries. In other words, 
all formulas given above hold in arbitrary dimension if 
we impose
\begin{equation}
\partial_\mu\xi^\mu_A=0.
\label{48}
\end{equation}
When (\ref{48}) holds,
the gauge transformations $\delta_\omega$ are
local Poincar\'e transformations. This raises
the question of whether there is a relation to
general relativity. 
The answer to this question is affirmative and easily obtained from the
following observation:
when (\ref{48}) holds, the ``Einstein-Hilbert action''
constructed
from the metric (\ref{22}),
\begin{equation}
S_{\mathrm{EH}}=-\frac 12\int d^nx\, \sqrt g R,
\label{49}
\end{equation}
is invariant under gauge transformations 
(\ref{18}) because equation (\ref{23})
reduces to a general coordinate transformation
of $g_{\mu\nu}$ with parameters $\varepsilon^\mu=\omega^A\xi^\mu_A$.
Now, consider the special case of an action given just by (\ref{49})
(without any additional terms), and assume
that $\{A^A_\mu\}$ contains (at least) the gauge
fields of all spacetime translations.
Then we may interpret (\ref{22}) as a field redefinition
which just substitutes new fields $g_{\mu\nu}$ for certain
combinations of the original field variables.
Since the action depends on the gauge fields only via the
new fields $g_{\mu\nu}$, it reproduces the standard
theory of pure gravitation as described by
general relativity. 

In fact, the argument is even 
more transparent when one works with the vielbein
(\ref{36}) rather than with the metric (\ref{22})
[according to (\ref{37}), the metric can be written
in terms of the vielbein, and thus action
(\ref{49}) can also be written in terms of the
vielbein, as usual].
That is, we may label the translations by
an index $\nu$ and choose the corresponding Killing vector
fields as $\xi^\mu_\nu=\delta_\nu^\mu$. Accordingly,
the gauge fields of translations are denoted by
$A_\mu^\nu$. (\ref{36}) may then be interpreted
as a field redefinition that substitutes
${e_\mu}^\nu$ for $A_\mu^\nu$. This field redefinition
is clearly local and invertible (at least locally), as
(\ref{36}) can obviously be solved for $A_\mu^\nu$ in terms
of ${e_\mu}^\nu$ and the gauge fields of Lorentz transformations.

The same argument
applies when we add to the integrand of
(\ref{49}) the first term of the matter
Lagrangian (\ref{33}) (the second term is not needed since
we consider only gauged Poincar\'e transformations here), the fermion
Lagrangian (\ref{41}) or
the Yang-Mills type Lagrangian (\ref{44}). Since these contributions
also depend on the gauge fields $A_\mu^A$ only via the ${e_\mu}^\nu$, 
the same field redefinition implies the equivalence to
general relativity coupled to
matter fields in the standard way.
\bigskip

{\em Acknowledgements.} The author is grateful to Sergei Kuzenko
for valuable discussions and for suggesting formula (\ref{33}),
and to Marc Henneaux for pointing out 
previous work on gauge theories of conformal symmetries.

\end{document}